# Spectroscopic evidence of intra-unit-cell charge redistribution in charge-neutral magnetic topological insulator Sb-doped MnBi$_6$Te$_{10}$

Khanh Duy Nguyen[1], Gabriele Berruto[1], Seng Huat Lee[2], Yunhe Bai[1], Haoran Lin[1], Qiang Gao[1], Zhiqiang Mao[2], Shuolong Yang[1*]
[1] Pritzker School of Molecular Engineering, the University of Chicago, IL 60637, USA
[2] Department of Physics, Pennsylvania State University, University Park, PA 16802, USA
*yangsl@uchicago.edu

The magnetic topological insulator MnBi$_6$Te$_{10}$ has emerged as a promising candidate for realizing the quantum anomalous Hall effect (QAHE), owing to its ability to retain ferromagnetism through precise control of anti-site defects. The next important task for realizing the QAHE is to tune the chemical potential into the energy gap formed by the broken time-reversal symmetry. Here we reveal an intra-unit-cell charge redistribution even when the overall doping suggests a near-charge-neutral condition. By performing time- and angle-resolved photoemission spectroscopy (trARPES) on the optimally 18% Sb-doped MnBi$_6$Te$_{10}$, we observe transient surface photovoltage (SPV) effects on both the MnBi$_2$Te$_4$ and single-Bi$_2$Te$_3$ terminations. Furthermore, we observe a time-dependent splitting of the band structure indicating multiple SPV shifts with different magnitudes. This observation suggests that adjacent plateaus with nominally the same terminating layer exhibit a strong intra-unit-cell charge redistribution, resulting in spontaneous electrical polarization. This is consistent with static micro-ARPES measurements revealing significant doping deviations from the charge-neutral configuration. Our findings underscore the challenges of engineering the family of Mn-Bi-Te materials to realize QAHE purely through chemical doping. Achieving the desired topological quantum phase requires both a uniform carrier doping and a ferromagnetic ground state. Furthermore, the light-induced polarization within each unit cell of ferromagnetic Mn(Bi$_{0.82}$Sb$_{0.18}$)$_6$Te$_{10}$ may open new possibilities for optoelectronic and spintronics.

**Introduction:**

Magnetic topological insulator MnBi$_2$Te$_4$ has been established as a platform for exotic quantum phenomena such as the quantum anomalous Hall effect (QAHE)[1] and the axion insulator state[2]. Related compounds in the MnBi$_{2n}$Te$_{3n+1}$ family have also garnered much research interest[1,3–6]. MnBi$_2$Te$_4$ and MnBi$_4$Te$_7$, both of which are A-type antiferromagnets, exhibit an even-odd-layer oscillation of the overall magnetization[6,7,4], leading to the switching between the QAHE and axion insulator states. This versatility is valuable for fundamental studies of novel topological quantum effects governed by symmetry and geometry[8–10]. Nonetheless, to realize a higher-temperature QAHE and explore its dissipationless edge states for quantum information applications, a stable ferromagnetic (FM) ground state in the MnBi$_{2n}$Te$_{3n+1}$ family is desired[11–13].

MnBi$_6$Te$_{10}$, a newer member of this family, has shown promise in stabilizing the FM ground state. Recent studies on its transport properties, electronic band structure, and atomic arrangement[5] highlight the critical role of anti-site defects in supporting the FM coupling between Mn sheets, separated by Bi-Te atomic layers[14]. Although this FM ground state is delicate, it can be finely controlled, opening a pathway toward realizing the QAHE at higher temperatures. When MnBi$_6$Te$_{10}$ is cleaved, three distinct surface terminations are exposed: MnBi$_2$Te$_4$ (MBT), single Bi$_2$Te$_3$ (1-BT), and double Bi$_2$Te$_3$ (2-BT). Previous ARPES studies[5] showed that the 2-BT termination is p-doped in the FM phase, while the MBT termination exhibits a sizable gap (~15 meV) at the Dirac point, potentially enabling exotic quantum phenomena when the Fermi level is tuned into this gap.

In many magnetically doped topological insulators, Sb doping is used to introduce holes and reach charge neutrality, thus enabling the QAHE[11,15]. However, its effectiveness in uniformly tuning the Fermi level of MnBi$_6$Te$_{10}$ remains unclear, especially considering that different terminations of MnBi$_6$Te$_{10}$ can exhibit drastically different dopings[5]. In this study, we report time- and angle-resolved photoemission spectroscopy (trARPES) measurements on the optimally 18% Sb-doped MnBi$_6$Te$_{10}$. Hall measurements indicate that the system is close to a state of overall charge neutrality. However, we observe transient surface photovoltage (SPV) effects on both the MBT and 1-BT terminations, indicating a higher hole doping in the bulk of the topological insulator[16–18]. Furthermore, we observe a time-dependent splitting of the band structure which indicates multiple SPV shifts with different magnitudes. This observation suggests that adjacent domains with a $c$-axis offset of one unit cell have a significant built-in electric potential difference. The potential difference indicates an intra-unit-cell charge redistribution, reconciling the termination-dependent doping revealed in static micro-ARPES measurements. Our work underscores the challenge of realizing QAHE through simple chemical doping and suggests that additional electron accumulation near the surface of ultrathin MnBi$_6$Te$_{10}$ can be harnessed for spintronic applications[19].

**Methods**:

**Sample growth**
The MnBi$_6$Te$_{10}$ single crystals were grown using a modified melt-growth method. A mixture of high-purity manganese powder (99.95%), bismuth shot (99.999%), antimony shot (99.999%), and tellurium ingot (99.9999+%) was mixed in a stoichiometric molar ratio of Mn:Bi:Sb:Te = 1:1.08:4.92:10, and sealed in an evacuated, carbon-coated quartz tube jacked by another evacuated quartz tube. The prepared ampoule was heated to 900 °C and held at this temperature for 12 hours to promote homogeneous melting. The mixture was then subjected to a controlled cooling and annealing process: it was cooled from 900 °C to 586 °C over 5 hours, from 586 °C to 576 °C at a rate of 0.1°C/hour, annealed at 576 °C for 72 hours, and finally quenched in water. Plate-like single crystals were obtained by cleaving along the basal plane of the resultant ingot. The quality and purity of the crystals were confirmed through X-ray diffraction (XRD) (Fig. S1).

**Transport measurements:**
The standard four-probe method was used for Hall measurements. Samples used for Hall resistivity $\rho_{yx}$ measurements had six attached leads, with one pair of current leads and two pairs of Hall voltage leads. The applied current was not aligned to any specific crystallographic axis in the $ab$-plane. Field sweeps of $\rho_{yx}$ were conducted for both positive and negative fields. The field dependence of $\rho_{yx}$ was obtained using $\rho_{yx} = [\rho_{yx}(+\mu_0 H) + \rho_{yx}(-\mu_0 H)]/2$. The transport mobility $\mu_H$ was estimated by $\mu_H = R_H/\rho_{xx}(0)$, in which $R_H$ is the Hall coefficient $R_H = d\rho_{yx}/d(\mu_0 H)$ and $\rho_{xx}(0)$ is the zero-field longitudinal resistivity. The carrier density for the samples was estimated via $n = 1/(eR_H)$, where $e$ is the electric charge.

**ARPES measurements**

The static- and tr-ARPES measurements were performed on the multi-resolution photoemission spectroscopy platform at the University of Chicago[20] at a base temperature of 8 K. The 6-eV laser for static ARPES was generated from a mode-locked Ti:sapphire oscillator with a repetition rate of 80 MHz. The trARPES setup featured a 200-kHz Yb:KGW laser accompanied by noncollinear optical parametric amplifiers (NOPAs). For the pump path, a two-stage NOPA was tuned to generate the signal beam at 800 nm. For the probe path, the 206 nm probe beam was obtained by frequency quadrupling the signal output of a one-stage NOPA. The energy resolutions of the static and trARPES setups were better than 4 and 20 meV, respectively. Focused probe beam waists, as characterized by the full width at half maximum, were $14 \times 17$ µm$^2$ and $24 \times 25$ µm$^2$ for the static and trARPES experiments, respectively. A systematic alignment procedure was adopted to ensure the overlap of the probed regions for static and trARPES[4]. Pump pulses were linearly polarized and 20-fs-long, with a beam waist of ~$72 \times 93$ µm$^2$ and a tunable incident fluence. The time resolution was determined to be ~150 fs, limited by the duration of the probe pulses. The MnBi$_6$Te$_{10}$ single crystals were cleaved in situ under a pressure $< 5 \times 10^{-11}$ mbar for the ARPES measurements.

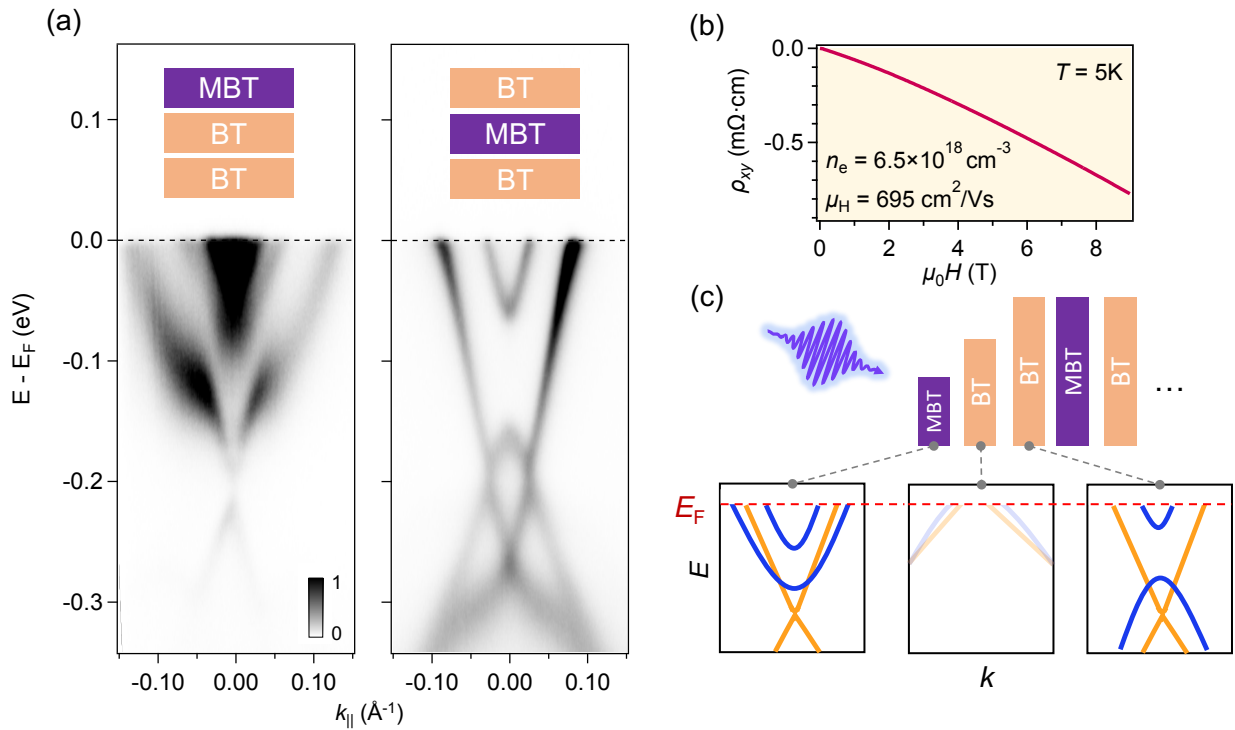

Fig. 1. Equilibrium band structure of Mn(Bi$_{1-x}$Sb$_x$)$_6$Te$_{10}$ with x = 0.18. (a) Band structures along the $\Gamma - M$ direction on the MBT and 1-BT terminations. (b) Hall measurements for 18% Sb-doped MnBi$_6$Te$_{10}$ at 5 K showing the carrier density ~ $6.5 \times 10^{18}$ cm$^{-3}$, near the charge-neutral point. (c) Illustrations of different cleaved terminations in MnBi$_6$Te$_{10}$ and sketches of their band structures.

**Results:**

In this experiment, we finely tune the doping level of Mn(Bi$_{1-x}$Sb$_x$)$_6$Te$_{10}$. A series of Hall measurements are performed on the samples with varying Sb concentration $x$, to determine the carrier types and densities[21]. At $x = 0.18$, the state of charge neutrality is confirmed through Hall measurements (Fig. 1(b)) with the total carrier density $n_e$ ~ $6.5 \times 10^{18}$ cm$^{-3}$, which is almost two order of magnitude smaller than $n_e$ ($x = 0$) = $1.98 \times 10^{20}$ cm$^{-3}$ of pristine MnBi$_6$Te$_{10}$ and $n_h$ ($x = 0.22$) = $2.37 \times 10^{20}$ cm$^{-3}$ of more hole-doped samples[21]. The FM ground state is also confirmed by an isothermal magnetization measurement (Fig. S2). At this doping level, the Fermi level is expected to reach the Dirac point, which is critical to driving the system into the QAH insulating phase. Surprisingly, the MBT and 1-BT terminations remain electron-doped. The Dirac point on the 1-BT termination in 18% Sb-doped MnBi$_6$Te$_{10}$ is shifted to -0.27 eV compared to -0.32 eV in undoped samples[5]. Meanwhile, the MBT termination gains additional electron charges as the Dirac point is shifted in the opposite direction, from -0.18 eV in undoped MnBi$_6$Te$_{10}$ to -0.22 eV in the 18% Sb-doped sample[5]. The electron-doped terminations reveal sharp topological surface states (TSSs). The MBT termination shows the hybridization between the TSS and Rashba states, which is commonly observed in the MBT family[4,5,13,22]. The 1-BT termination features gapless TSSs along with a V-shaped conduction band and a Λ-shaped valence band. Therefore, the transport and ARPES data present no fundamental changes in the magnetic and topological properties as the system is doped with Sb. Nevertheless, due to the conservation of the total charge carrier density within the crystal, the 2-BT termination is expected to be hole-doped. Despite extensive efforts, we were unable to identify the 2-BT termination in our experiments. This contrasts with the undoped FM MnBi$_6$Te$_{10}$ system, where all three terminations were readily identifiable[5,23]. In the undoped material, the 2-BT band structure differs from that of the 1-BT mainly due to a significant downward shift in the chemical potential. In our 18% doped system, shifting the chemical potential of the 1-BT band structure to more than 0.15 eV below the Dirac point results in a diffusive ARPES spectrum (Fig. S3). This spectrum closely resembles that obtained from poorly cleaved surfaces, making it difficult to distinguish the 2-BT termination. The unexpected electron doping in the ARPES spectra for the MBT and 1-BT terminations, deviating from what transport measurements suggest, may imply charge redistribution within the crystal, which will be discussed in detail later.

We then conducted trARPES experiments on the MBT and 1-BT terminations to investigate band-bending effects. Figure 2(a) shows the surface photovoltage (SPV)[18,24] on the MBT termination at representative delays. At negative time, photoelectrons are generated before the pump pulse reaches the sample surface. When the pump pulse arrives, although the photoelectrons have escaped the surface, they remain close to the sample surface to experience the electric field created by the photoexcited carriers. This is the reason that the SPV-induced energy shift extends to 30 ps toward the negative delays (Fig. 2(c)), with the time scale determined by the spatial profile of the pump beam[25,26]. This phenomenon, well-documented in trARPES experiments on GaAs[25] and recently on topological insulators like Bi$_2$Se$_3$ and Bi$_2$Te$_3$[17,27–29], is spectroscopically manifested by the band shift near time zero, as shown in Fig. 2 (a). In these cases, the bulk remains p-type, but the surface is metallic due to the presence of the TSSs, causing a downward band bending at the surface. When the system is pumped by infrared light, electron-hole pairs are generated, and the migration of holes and electrons in opposite directions flattens the bands, causing the Fermi level at the surface to rise. In our system, we observed a maximal band shift of ~50 meV at time zero. This energy shift is saturated when the pump fluence is higher than 4.8 μJ/cm$^2$, as confirmed by varying the fluence from 0.25 to 95 μJ/cm$^2$ (Fig. S4). The positive SPV observed in 18% Sb-doped MnBi$_6$Te$_{10}$ verifies the downward band bending near the surface, indicating a metallic surface and a relatively more hole-doped bulk[5,13,30], similar to the case of the p-type Bi$_2$Se$_3$ and Bi$_2$Te$_3$ crystals[17]. The band bending in Sb-doped MnBi$_6$Te$_{10}$ may differ fundamentally from other homogeneous topological insulators due to non-uniform charge distribution within the bulk. The different carrier dopings of different layers in one unit cell may lead to an intra-unit-cell band bending in addition to the overall trend (Fig. 3(b)). However, the SPV effect presented in Fig. 2(a) does not allow us to discern the intra-unit-cell band bending.

A dual-SPV effect is observed at the 1-BT termination, as shown in Fig. 3, which manifests the strong intra-unit-cell charge redistribution. The transient Fermi level shift is significantly larger than that in the MBT termination using the same incident fluence of 95 μJ/cm$^2$, with a maximum shift of approximately 100 meV, comparable to bulk-insulating $Bi_2Se_3$ and $Bi_2Te_3$ crystals[17,27,28]. Intriguingly, trARPES experiments reveal the splitting of the band structure into two replicas. These two replicas display different time-dependent SPV shifts. The maximum splitting occurs near time zero, with an energy splitting of ~90 meV and a momentum splitting of ~0.01 Å$^{-1}$. This effect is not present in static ARPES, which qualitatively shows a single set of band structure [Fig. 3(a)]. In the static ARPES spectrum, we fit the momentum distribution curve (MDC) near the Fermi crossing of the TSS (peak S) using a Lorentzian function and obtain a momentum linewidth of 0.028 Å$^{-1}$. A similar analysis for the split TSSs at -1 ps yields momentum linewidths of 0.023 and 0.026 Å$^{-1}$ for the upper ($S_1$) and lower ($S_2$) TSSs, respectively. The slightly broader linewidth in the static ARPES data indicates that the signal arises from the superposition of multiple domains with slight angular misalignments. Given the charge redistribution in Sb-doped $MnBi_6Te_{10}$, we propose the following explanation for the dual-SPV effect. Different layers within a unit cell have distinct charge carrier densities, but the overall system remains charge-neutral, leading to an effective p-n junction within each $MnBi_6Te_{10}$ unit cell, as illustrated in Fig. 3(b). Upon cleaving the crystal, adjacent 1-BT domains with a *c*-axis offset of multiple integers of unit cells can be obtained and probed by the same 6-eV beam. At equilibrium, these surfaces are electrically connected via the topological surface state, producing the same Fermi level observed in static ARPES. However, when SPV occurs, the top 1-BT termination is influenced by an additional dipole field due to electron-hole separation in the top unit cell, resulting in two distinct SPV effects. Furthermore, photoelectrons emitted near domain boundaries may experience a lateral electric field due to in-plane non-uniform potential, leading to the momentum splitting.

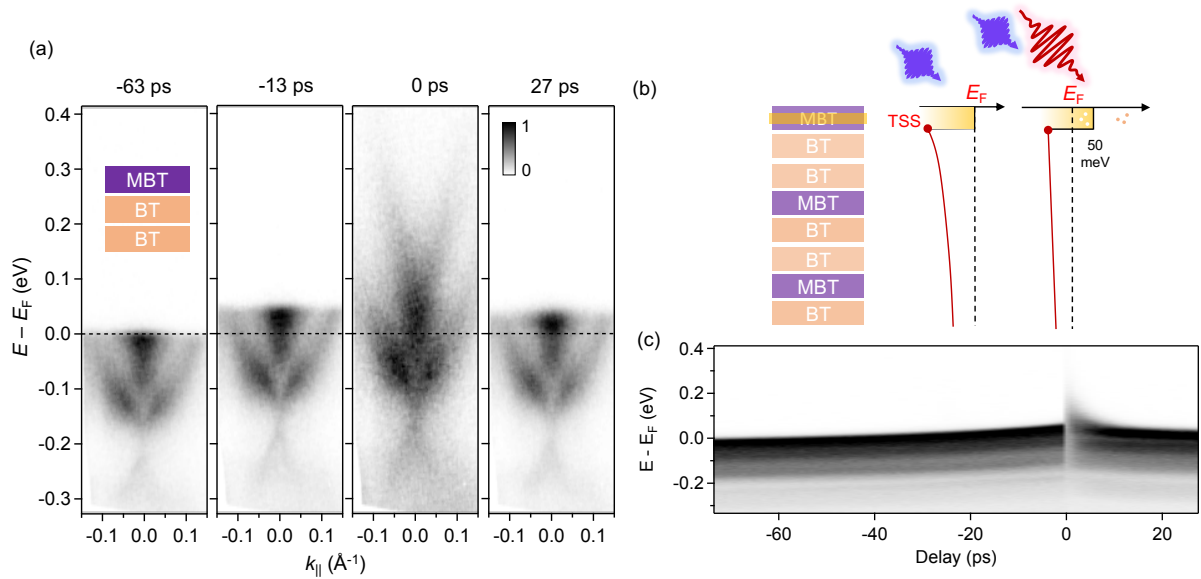

Fig. 2. Surface photovoltage effect on the MBT termination. (a) trARPES spectra on the MBT termination at representative delays. The incident IR pump fluence is 95 μJ/cm$^2$, at the over-saturated limit. (b) Proposed charge redistribution in $Mn(Bi_{0.82}Sb_{0.18})_6Te_{10}$ and band bending toward the surface. When the sample is excited by the IR pump, electron-hole pairs are formed, flattening the bands and causing the uplifting of the whole band structure. The red solid line represents the overall band bending. The black dashed line stands for the equilibrium Fermi level. (c) Time-dependent energy distribution curves (EDC) around the *Γ* point. The zero energy corresponds to the Fermi level at equilibrium.

We note that the trARPES experiment alone cannot determine the vertical offset between the two 1-BT domains to be exactly one unit cell. However, when cleaving a MnBi$_6$Te$_{10}$ crystal, most step edges are associated with either an MBT layer (1.3 nm height) or a BT layer (1 nm height)[31]. Our high-quality band structure on the 1-BT termination in Fig. 3(a) suggests that no other terminations contribute to the ARPES spectrum. Hence, the adjacent 1-BT domains must be separated by a sharp edge with the vertical offset of multiple unit cells. Notably, a step edge with the offset of 1 unit cell, which consists of 1 MBT layer and 2 BT layers, is already rare. A bunched step edge with the height of several unit cells is extremely difficult to obtain. Therefore, we attribute our observed dual-SPV effect to adjacent 1-BT domains with only one-unit-cell vertical offset. We emphasize that such multi-SPV effects are not restrained to the 1-BT termination. We have observed a triple-SPV on an MBT termination indicating the presence of three adjacent MBT domains (Fig. S5).

The multi-SPV effects directly suggest the intra-unit-cell charge redistribution. Fig. 3(a) illustrates that the SPV shift due to the electron-hole separation in the top unit cell is ~90 meV, which accounts for the majority of the overall SPV shift when referenced to the equilibrium band structure.

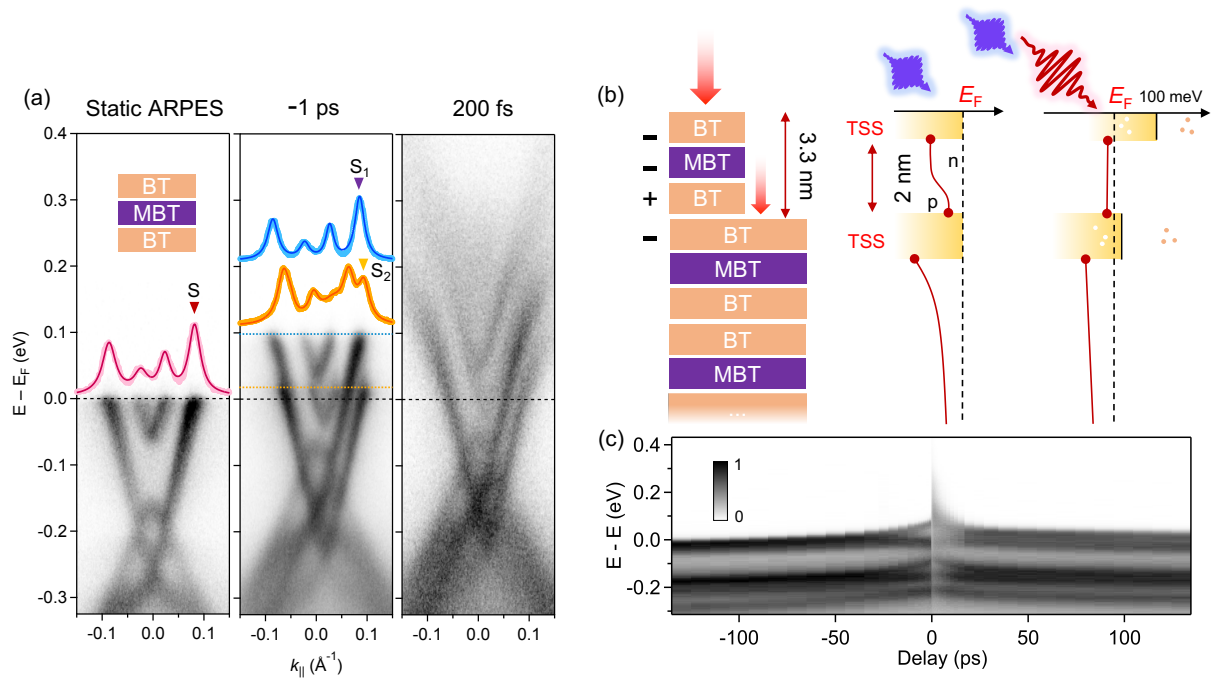

Fig. 3. Multiple surface photovoltages on the 1-BT termination. (a) Static ARPES spectrum and trARPES spectra on the 1-BT termination with momentum distribution curves (MDCs) taken at the equilibrium Fermi level (static ARPES) and the shifted Fermi levels of two replica bands at -1 ps. Peak S at equilibrium is split into peaks $S_1$ and $S_2$ at -1 ps. The time-dependent spectra were obtained using an IR pump fluence of 95 μJ/cm$^2$. (b) Proposed charge redistribution in Mn(Bi$_{0.82}$Sb$_{0.18}$)$_6$Te$_{10}$ and band bending toward the surface. The cartoons illustrate the dual-SPV effect at 2 adjacent exposed 1-BT terminations. The red solid line represents the overall band bending. The black dashed line stands for the equilibrium Fermi level. (c) Time-dependent EDCs around the $\Gamma$ point clearly show the band splitting starting from around -30 ps and lasting more than 100 ps after time zero.

**Discussion**:

Unlike conventional semiconductors where doping can be homogeneously controlled, the family of MBT-derived materials are quasi-2D with heterolayers. Especially in $MnBi_6Te_{10}$, uniform doping control is shown to be challenging. Despite that the transport data indicates a global charge-neutral regime, ARPES and trARPES reveal a different picture. Static micro-ARPES gives distinct band structures at different terminations, each exhibiting a unique doping level. Although the 2-BT termination is not directly observed, the overall charge neutrality suggests that the 2-BT layer must be p-doped, which is consistent with our expectation based on previous studies[5]. The electron-doping in MBT and one of the BT layers, together with the hole-doping in the other BT layer, forms an intrinsic n-p junction. This effect explains the dual-SPV or triple-SPV (Fig. S5) effects observed near the boundaries of adjacent domains which are separated by a vertical offset of one unit cell.

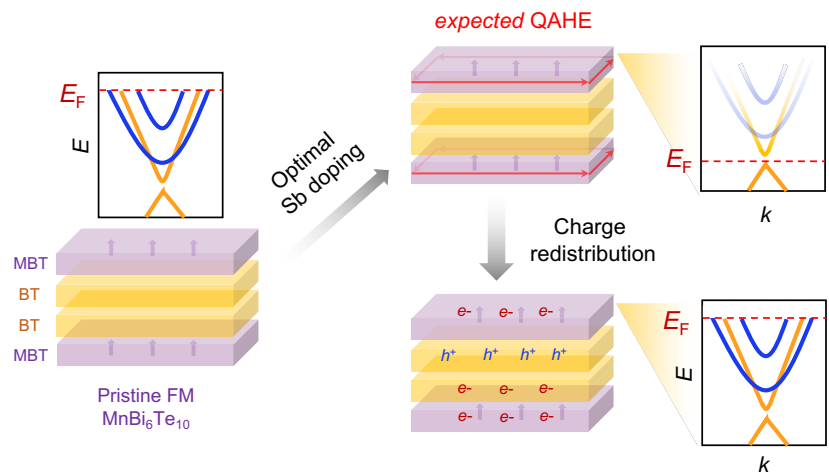

Fig. 4. Charge redistribution in the charge neutral $Mn(Bi_{0.82}Sb_{0.18})_6Te_{10}$ prevents the material from realizing the quantum anomalous Hall effect. Instead of the homogenous charge neutrality, the imbalance of the charge distribution always leaves the MBT layer electron-doped.

The charge redistribution in nominally charge-neutral $MnBi_6Te_{10}$ hinders the realization of exotic quantum phenomena such as the QAHE (Fig. 4). Our experimental results suggest that more engineering is needed beyond chemical doping to realize these charge-neutrality-based transport effects. We attribute the charge redistribution to a high level of Mn-Bi/Sb anti-site defects, which is exploited to stabilize the FM phase[5,14]. In undoped $MnBi_6Te_{10}$, previous literature showed that Mn-Bi mixing and Mn vacancies stabilize the FM phase[5]. Notably, Sb doping in $MnBi_{2n}Te_{3n+1}$ systems promotes more anti-site defects[32]. This is due to the smaller size of Sb atoms compared to Bi, leading to an easier mixing with Mn[32]. Substituting $Mn^{2+}$ with $Bi^{3+}$ or $Sb^{3+}$ in the MBT layer leads to electron doping; the inverse substitution in the BT layers in general leads to hole doping. In reality, the two BT layers in one unit cell are electron- and hole-doped, respectively. By comparing with the existing literature on $MnBi_{2n}Te_{3n+1}$ systems[5,23], we adopted the conventional assignment that the more electron-doped BT termination is the 1-BT termination. However, we note that there is no reason to observe a global doping asymmetry between the two BT layers, as the "top" and "bottom" BT layers in one unit cell can be swapped by simply flipping the crystal. Hence we believe that the doping asymmetry between the top and bottom BT layers, as shown in Fig. 4, occurs stochastically. Our earlier ARPES data on pristine AFM $MnBi_6Te_{10}$ revealed more uniform doping across different layers[5], with electron-doping at all terminations, in contrast to the strongly non-uniform doping in the FM phase. Thus, an integrated approach combining Sb doping and the control of Mn-Bi/Sb anti-site defects could improve the uniformity of carrier densities in the FM phase. Nonetheless, the light-induced charge polarization in $MnBi_6Te_{10}$ may open new possibilities of applications in optoelectronics and energy harvesting. For instance, SPV in topological insulators with a strong spin-orbit coupling can boost ultrafast spin transport in spintronics[19] with the robustness of topological protection[33,34].

**Conclusion:**

In summary, we investigated the electronic structure of 18% Sb-doped FM $MnBi_6Te_{10}$. The transport data from Hall measurements indicate that the overall charge neutrality is achieved. However, micro-ARPES reveals that the MBT and 1-BT terminations remain electron doped. The trARPES measurements show consistent SPV effects. In particular, the intra-unit-cell charge migration in Sb-doped $MnBi_6Te_{10}$ induces electrical polarization, manifested as multiple SPV effects in which bands are split into several identical replicas, translated in both energy and momentum. Our findings highlight the challenges in engineering Mn-Bi-Te materials to realize the QAHE solely through chemical doping. Achieving the desired topological quantum phase requires both a uniform carrier doping and a ferromagnetic ground state. On the other hand, the polarization within each unit cell of ferromagnetic $MnBi_6Te_{10}$ may create new possibilities for optoelectronic and spintronic devices. For example, Sb-doped $MnBi_6Te_{10}$ can be a strong material platform for the next-generation magnetic tunnel junction devices, which requires both highly efficient charge-to-spin conversion[35] as well as electric-field control[36].

**Acknowledgement:** The time-resolved ARPES work was supported by The U.S. Department of Energy (grant no. DE-SC0022960). Support for crystal growth and characterization at Penn State was provided by the National Science Foundation through the Penn State 2D Crystal Consortium-Materials Innovation Platform (2DCC-MIP) under NSF Cooperative Agreement DMR 2039351.

**Data Availability Statement:** The data supporting this article have been included in the main text or the Supplementary Information.

# Supplementary Information

**Spectroscopic evidence of intra-unit-cell charge redistribution in charge-neutral magnetic topological insulator Sb-doped MnBi$_6$Te$_{10}$**

Khanh Duy Nguyen[1], Gabriele Berruto[1], Seng Huat Lee[2], Yunhe Bai[1], Haoran Lin[1], Qiang Gao[1], Zhiqiang Mao[2], Shuolong Yang[1*]
[1] Pritzker School of Molecular Engineering, the University of Chicago, IL 60637, USA
[2] Department of Physics, Pennsylvania State University, University Park, PA 16802, USA

*yangsl@uchicago.edu

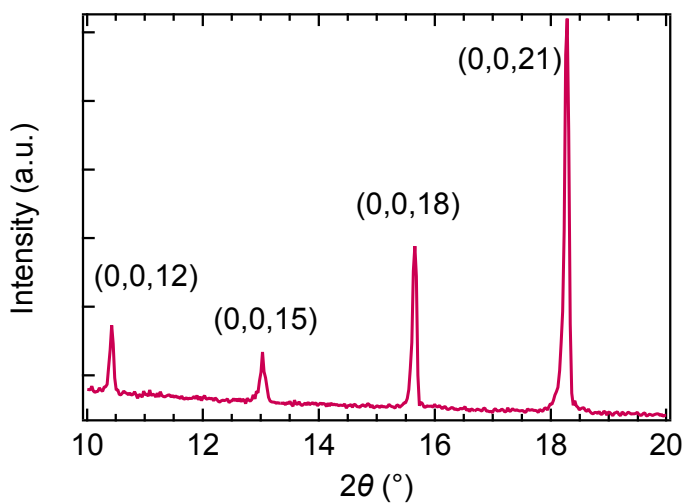

**Fig. S1. X-ray diffraction pattern of Mn(Bi$_{1-x}$Sb$_x$)$_6$Te$_{10}$ crystal, with $x$ = 0.18**. The four peaks are at almost exactly the same locations as in undoped MnBi$_6$Te$_{10}$, indicating the same crystal structure. The wavelength of the X-ray source is 1.54059Å.

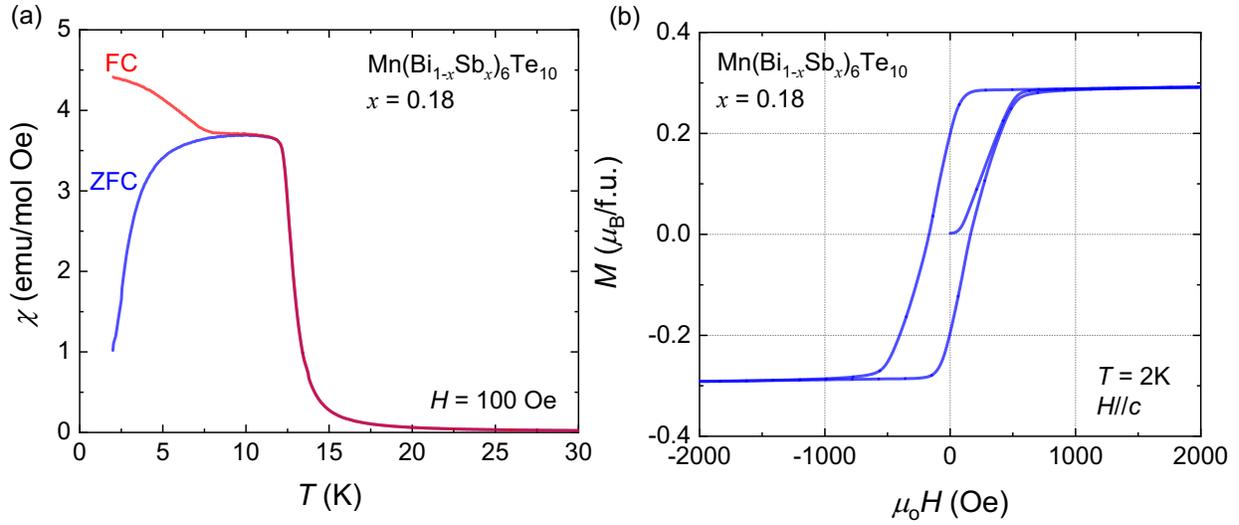

**Fig. S2. Ferromagnetic ground state of Mn(Bi$_{1-x}$Sb$_x$)$_6$Te$_{10}$ crystal, with *x* = 0.18.** (a) Temperature dependent zero-field-cooled (ZFC) and field-cooled (FC) magnetic susceptibilities indicate T$_C$ ~ 13 K. (b) Isothermal magnetization curves with the magnetic field applied along the *c*-axis taken at 2 K.

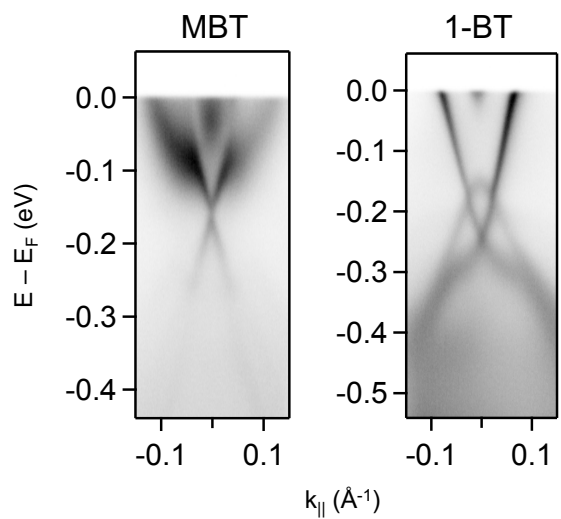

**Fig. S3. Band structure of the MBT and 1-BT termination with wider ranges of binding energy.** Both terminations show that the spectral features become diffuse at more than 0.15 eV below the Dirac point. This characteristic leads to the consequence that if the 2-BT termination is heavily hole-doped, the corresponding spectrum will show diffusive features, which cannot be distinguished from spectra originating from poorly cleaved areas.

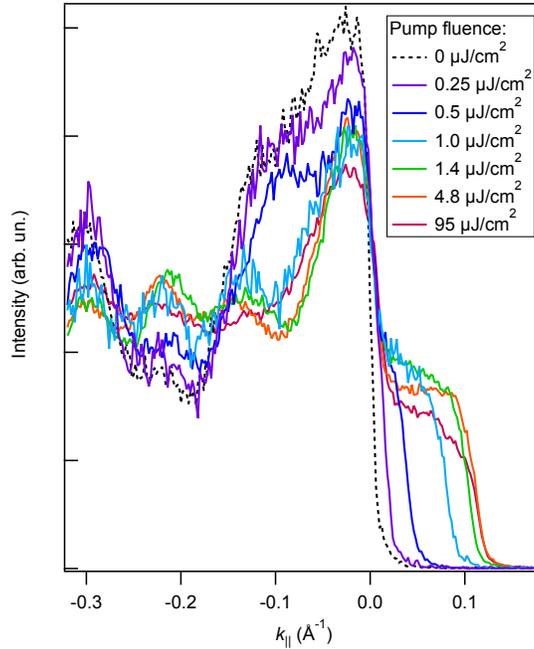

**Fig. S4. Fluence-dependent surface photovoltage in the MBT termination**. The figure shows energy distribution curves (EDCs) taken around the Γ point at -1 ps for different pump fluences.

A series of time-resolved ARPES (tr-ARPES) experiments were conducted on the MBT termination, as shown in Fig. 2 of the main text. The crystal was excited using 1.5 eV light with incident fluences varying from 0.25 to 95 µJ/cm². As illustrated in Fig. S4, the SPV effect, indicated by an energy shift, saturates at approximately 100 meV with a pump fluence of 4.8 µJ/cm². Notably, even when the fluence was significantly increased to 95 µJ/cm², no further shift was observed. This saturation behavior can only be attributed to the SPV effect in the "flat-band condition" rather than any conventional charging effect induced by photoexcitation.

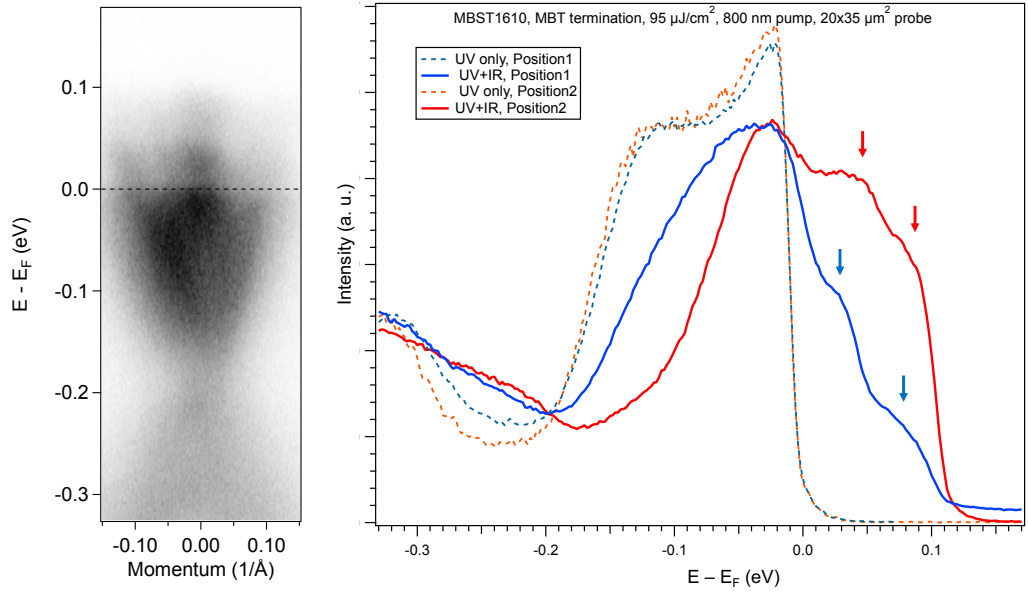

**Fig. S5: Triple SPV effect on a region of the MBT termination**. Left) The trARPES spectrum taken at –1 ps shows three replicas of the band structure. Right) EDCs taken around the Γ point at –1 ps for two positions where the triple SPV effect was observed. The experiment was done with an incident pump fluence of 95 μJ/cm$^2$.

On a different cleaved surface with an MBT-like termination, a triple SPV effect is observed, as evidenced by the splitting of the band structure into three nearly equidistant copies in energy, as shown in Fig. S5 (left). Energy distribution curves (EDCs) taken around the Γ point provide further quantitative insight into this SPV effect, as depicted in Fig. S5 (right). The energy spacing between closest band replicas is ~50 meV. This observation is consistent with the assumption that one-unit-cell offset in the *c*-direction between adjacent crystal domains induces a fixed potential difference across these surfaces.